\documentclass{aa}  

\usepackage{graphicx}
\usepackage{txfonts}
\usepackage{empheq}
\usepackage{threeparttable} 
\usepackage{amsmath}
\usepackage{siunitx}
\usepackage{comment}
\usepackage{lineno}

\begin{document} 


\title{Noble gas depletion on Titan: Clathrate sequestration during the open ocean phase}
\authorrunning{Amsler Moulanier et al.}
\titlerunning{Noble gas depletion on Titan}

   \author{A. Amsler Moulanier
          \inst{1},
          O. Mousis\inst{1,2},
          A. Bouquet\inst{3,1}
          \and
          N. H. D. Trinh\inst{4,1}
          }

   \institute{Aix-Marseille Université, CNRS, CNES, Institut Origines, LAM, Marseille, France\\
   \email{alizee.amsler@lam.fr}
         \and
             Institut Universitaire de France (IUF)
         \and
            Aix-Marseille Université, CNRS, Institut Origines, PIIM, Marseille, France
        \and
            University of Science and Technology of Hanoi (USTH), Hanoi, Vietnam\\
             }

\date{Received March 6, 2025; accepted April 29, 2025}

  \abstract
{A plausible explanation for the absence of primordial argon, krypton, and xenon in Titan's current atmosphere is that these gases were sequestered in clathrate hydrates during Titan's "open-ocean" phase.
We examine how clathrate hydrate formation at Titan's ocean surface in its early history may have contributed to noble gas depletion in the primordial atmosphere.
Starting with vapor-liquid equilibrium modeling between water and volatiles, we used a statistical thermodynamic model to determine the clathrate hydrate crust thickness needed to deplete the primordial atmosphere of noble gases.
Our computations suggest that if Titan's volatile budget was delivered by icy planetesimals with a comet-like composition, its primordial atmosphere should be rich in CO$_2$ and CH$_4$, with NH$_3$ largely retained in water as ions. We show that at 273.15 K, a clathrate crust tens of kilometers thick would deplete the primordial atmosphere of xenon and krypton.
The lack of primordial argon in Titan's atmosphere may result from the partial de-volatilization of its accreted materials.}


\maketitle
%

\section{Introduction}\label{sec:intro}

A notable feature of Titan's atmosphere is the absence of heavy noble gases except argon, as revealed by the Gas Chromatograph Mass Spectrometer (GCMS) aboard the Huygens probe during its descent to the surface in January 2005. The GCMS detected argon in two forms: primordial $^{36}$Ar and radiogenic $^{40}$Ar, a decay product of $^{40}$K. However, other primordial noble gases, including $^{38}$Ar, krypton (Kr), and xenon (Xe), were not detected, with GCMS placing upper limits on their atmospheric mole fractions of $<10^{-8}$ \citep{niemann_abundances_2005,niemann_composition_2010}.

Titan is believed to have formed from a mixture of ices, silicates, and organic materials \citep{muller-wodarg_origin_2014,reynard_carbon-rich_2023}. A plausible origin for Titan's volatile phase is that its building blocks had a comet-like volatile composition \citep{alibert_formation_2007,mousis_primordial_2009}. This is supported by the D/H ratio in Enceladus's water vapor matching the value observed in comets \citep{waite_liquid_2009}, which suggests similar building blocks for Titan and Enceladus \citep{mousis_primordial_2009, mousis_clathration_2009, mousis_determination_2009}. Titan's noble gas depletion could be explained by the trapping of Ar, Kr, and Xe in clathrate hydrates (hereafter clathrates) within its icy shell \citep{osegovic_compound_2005, thomas_theoretical_2008, mousis_removal_2011}, with sequestration occurring at the surface and facilitated by cryovolcanism \citep{mousis_removal_2011}.

However, early in Titan's history, high surface temperatures likely supported a global ocean in contact with its primordial atmosphere \citep{lunine_clathrate_1987, lunine_massive_1992}, enabling clathrate formation at the ocean--atmosphere interface. This study explores how Titan's hydrosphere acted as a clathrate reservoir. The compositions of Titan's ocean and atmosphere depended on the volatile inventory of its building blocks, with the CO$_2$/NH$_3$ mass ratio influencing atmospheric composition and thickness \citep{marounina_role_2018, amsler_moulanier_role_2025}. Assuming a comet-like volatile inventory, the early atmosphere likely contained CH$_4$ and CO$_2$ \citep{lunine_origin_2009, muller-wodarg_origin_2014}, with NH$_3$ as a key hydrospheric volatile \citep{lunine_clathrate_1987, grasset_cooling_1996, tobie_titans_2012}. NH$_3$ likely contributed to Titan's current N$_2$ atmosphere \citep{atreya_evolution_1978, jones_estimated_1987, mckay_high-temperature_1988, lunine_present_1989, glein_absence_2009} and may have acted as an antifreeze \citep{lunine_clathrate_1987, grasset_cooling_1996}.

In this study we first analyzed the evolution of Titan's primordial hydrosphere composition, assuming its volatile inventory was primarily delivered by icy building blocks. Second, we applied a statistical thermodynamic model to investigate whether clathrate formation at the surface of Titan's ocean explains the depletion of primordial noble gases and its impact as well as the impact of clathrate formation on atmospheric composition.

\section{Methods}

We aimed to calculate the minimum thickness of the mixed clathrate crust formed during Titan's primordial phase to explain the current noble gas depletion. The model used in this study simulates the evolution of the composition of Titan's primordial hydrosphere as temperature decreases, leading to the formation of an ice crust. If the stability conditions for clathrate formation are met, the model calculates how this process could affect the distribution of volatiles in Titan's primordial atmosphere.

 We assumed that shortly after accretion, Titan's surface temperature was high enough to support a primordial ocean in direct contact with the atmosphere \citep{lunine_clathrate_1987,lunine_massive_1992}. These two thermodynamic phases were assumed to be in equilibrium, and the partitioning of volatiles between them was determined accordingly. The model considers both the liquid-vapor equilibrium and the chemical equilibrium between CO$_2$ and NH$_3$ within the ocean. Starting from a given bulk composition, surface temperature, and an initial guess for the atmospheric composition, we calculated the equilibrium surface pressure of the global ocean and the distribution of volatiles between the primordial atmosphere and the ocean. Once equilibrium is reached, a mass balance check ensures consistency with the initial volatile mass in the bulk composition. If convergence is not achieved, the process is repeated, adjusting the atmospheric partial pressures until the volatile partitioning between the atmosphere and ocean satisfies bulk mass conservation. For a detailed formulation of the equations governing this equilibrium, we refer to Appendix \ref{sec:L-V} and \cite{amsler_moulanier_role_2025}. The model does not account for water--rock interactions between the rocky mantle and the ocean. 
 If the total atmospheric pressure at the surface exceeds the dissociation pressure of the clathrates (Eq. \ref{eq:P_diss}), mixed clathrates can form at the ocean surface. The dissociation pressure of the mixed clathrate is given as

\begin{equation}  
    (P_{diss}^{mix})^{-1} = \sum_{i=1}^{N}\frac{y_i}{P_{diss}^i},
    \label{eq:P_diss}  
\end{equation}  

\noindent where $P_{diss}^i$ is the dissociation pressure of the pure clathrate of gas $i$, $y_i$ is the atmospheric mole fraction of gas $i$, and $P_{diss}^{mix}$ is the dissociation pressure of the mixed clathrate \citep{lipenkov_stability_2001}. When the stability conditions for clathrates are met, their composition is determined using a statistical thermodynamic formation model derived from \cite{mousis_abundances_2013} and detailed in Appendix \ref{sec:app_clath}. 

 For a given pressure, temperature and gas composition, our model calculates the fractions of volatiles trapped in mixed clathrates. Since in our model CH$_4$ and CO$_2$ are the most abundant gases in the atmosphere, they are the primary clathrate formers, and their predominant clathrate structure is expected to be structure I \citep{sloan_clathrate_2008}. Consequently, all calculations were performed for structure I clathrates (SI).  At each time step, the thickness of the mixed clathrate layer is determined based on a convergence criterion. While clathrate formation is generally considered to be a rapid process \citep{englezos_kinetics_1987, uchida_twostep_2004}, its kinetics remain poorly constrained. Consequently, interpretation of the time step in terms of years is challenging, and the results are instead expressed as a function of clathrate crust thickness. We stress that, as observed on Earth, the kinetics of clathrate formation occurs on timescales between 1,000 and 100,000 years. However, while this timescale is relevant in Earth's context, it still needs to be evaluated for Titan. Then, as the crust thickens, the diffusion of the gas becomes less efficient, slowing growth \citep{ruppel_timescales_2020}.

 An SI clathrate consists of eight cages per unit cell, with a volume of $1.73\times 10^{-27}$ m$^3$ \citep{sloan_clathrate_2008}. In our case, mixed clathrates form, with each cage accommodating one gas molecule. The amount of mixed clathrates required to trap noble gases is expressed as the corresponding trapped gas volume per kilometer of crust. This volume is then subtracted from the primordial atmosphere, which serves as the gas reservoir for clathrate formation. Finally, the density of the formed clathrates is computed using the method outlined in Appendix \ref{sec:density}. We emphasize that the feedback effects on ocean composition due to changes in atmospheric composition were not calculated in this study.

\section{Results}\label{sec:Results}

We investigated the composition of Titan's primordial hydrosphere during two distinct phases: the open-ocean phase, when the ocean directly interacted with the atmosphere, and the closed-ocean phase, when an icy crust formed, isolating the ocean from atmospheric exchange.

\subsection{Combined evolution of Titan's ocean and atmosphere compositions during the open-ocean phase}

The composition and evolution of Titan's primordial hydrosphere were examined based on a specific bulk composition, with particular emphasis on the role of NH$_3$, whose abundance ranges from 1.5 to 15 wt$\%$ in the ocean \citep{tobie_titans_2012}. The dissolution of ammonia and carbon dioxide in the global ocean can significantly influence atmospheric composition and thickness through the chemical balance between NH$_3$ and CO$_2$ speciation products. In particular, the ratio of dissolved CO$_2$ to NH$_3$ is a key factor in determining whether a CO$_2$-rich atmosphere can be sustained \citep{marounina_role_2018,amsler_moulanier_role_2025}.  

\begin{figure}[htpb]
\centering
\includegraphics[width=8.7cm]{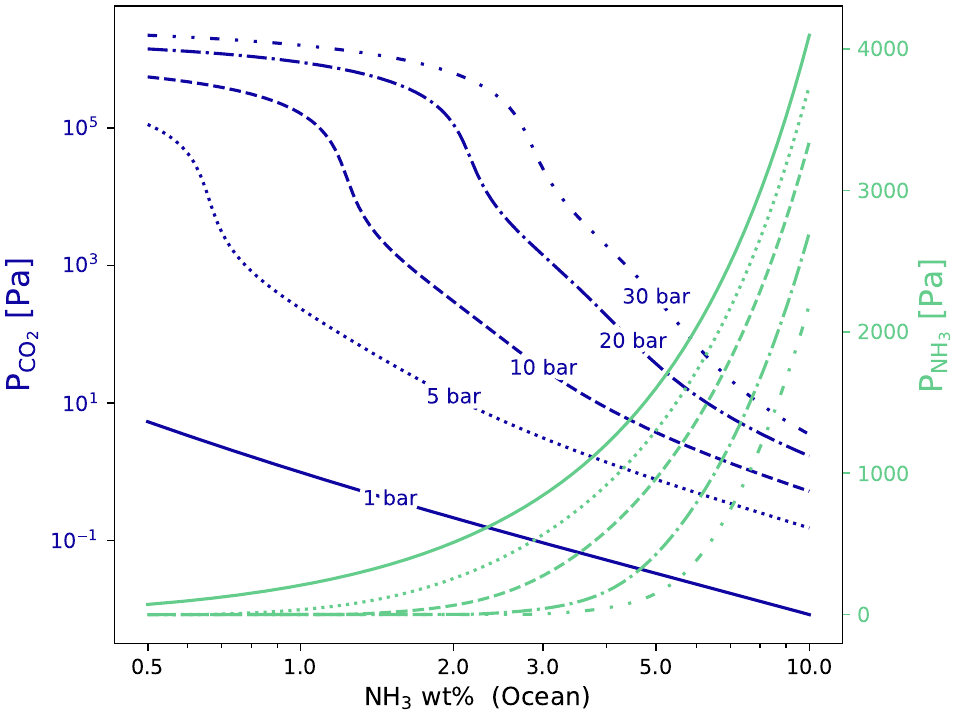}
\caption{Equilibrium partial pressures of CO$_2$ (dark blue) and NH$_3$ (green) in Titan's primordial atmosphere as a function of the total NH$_3$ dissolved in the ocean. Each set of green and blue curves, distinguished by different line styles, represents a specified initial CO$_2$ atmospheric pressure prior to chemical equilibrium between the ocean and atmosphere. The initial theoretical $P_{CO_2}$ values of 1, 5, 10, 20, and 30 bar correspond to CO$_2$ bulk masses of $7.8\times10^{19}$ kg, $3.79\times10^{20}$ kg , $7.38\times10^{20}$ kg, $1.37\times10^{20}$ kg, and $1.92\times10^{30}$ kg, and a dissolved CO$_2$ mole fraction of $1.36\times10^{-3}$, $6.78\times10^{-3}$, $1.36\times10^{-2}$, $2.71\times10^{-2}$, and $4.07\times10^{-2}$, respectively.}
\label{fig:PCO2vsPNH3}
\end{figure}

 Figure \ref{fig:PCO2vsPNH3} presents the equilibrium partial pressures of CO$_2$ and NH$_3$ in Titan's primordial atmosphere at 273.15 K, as a function of the total NH$_3$ fraction in the ocean and the theoretical initial CO$_2$ partial pressure. The results indicate that when the NH$_3$ concentration exceeds 2\%, even an initially high atmospheric CO$_2$ abundance (>10 bar) cannot be sustained above Titan's primordial ocean. This finding challenges the hypothesis that Titan once hosted a thick CH$_4$- and CO$_2$--rich primordial atmosphere \citep{lunine_origin_2009,muller-wodarg_origin_2014} in direct contact with an NH$_3$-rich ocean \citep{lunine_clathrate_1987,tobie_titans_2012}. It also raises questions about the feasibility of surface clathrate formation during the open-ocean phase, since this process requires the total atmospheric pressure to exceed the clathrate dissociation pressure. As shown by \cite{marounina_role_2018}, we suggest that, in the temperature range considered, a thick CO$_2$-rich primordial atmosphere in equilibrium with a NH$_3$-rich ocean would not have been sustainable.

 The composition of Titan's primordial atmosphere, and its relative tendency to favor CO$_2$ or NH$_3$, is inferred from its primordial bulk composition. Here, Titan's volatile inventory is assumed to have originated from volatile-rich icy planetesimals with a comet--like composition, with the volatile fraction in the ice phase analogous to that measured in comet 67P/Churyumov-Gerasimenko (67P/C-G) \citep{rubin_elemental_2019}.
Given Titan's bulk density, the total mass of ice in its composition is assumed to be $\sim$50\% \citep{lunine_origin_2009, tobie_titans_2012}. In this study, we assumed that about 30\% of Titan's mass contributed to the melting of ice that formed the primordial atmosphere and ocean \citep{lunine_origin_2009}, corresponding to a melted ice fraction of 60\%. By computing the normalized volatile mass distribution in ice, we could estimate the total bulk mass of each volatile delivered to Titan's hydrosphere (see Table \ref{tab:hydrosphere_compo_273}). Based on this volatile inventory, Titan's primordial ocean would be approximately 500 km deep. If the melted ice fraction were to reach 80\% \citep{monteux_can_2014}, the partial pressures above the ocean would increase. Specifically, higher CO$_2$ pressures could support the existence of liquid CO$_2$. However, a higher surface pressure could promote clathrate formation at a temperature slightly higher than 282 K if 80\% of the ice melts.

Using our model (Appendix \ref{sec:L-V}), we computed the thermodynamic equilibrium between the primordial atmosphere ocean and how volatiles are partitioned between the two phases. Assuming comet-like building blocks, a significant amount of CO$_2$ is released from the ice (4.12$\times 10^{21}$ kg). Our computations show that such a mass of CO$_2$ introduced into the hydrosphere would lead to a considerable pressure at Titan's surface. To model a physically accurate system, we capped the initial partial pressure of CO$_2$ at its saturation pressure as computed with Antoine's law. As mentioned previously, both CO$_2$ and NH$_3$ should be present within the primordial hydrosphere of Titan in significant proportion. Although NH$_3$ has a higher solubility than CO$_2$, Table \ref{tab:hydrosphere_compo_273} shows that at 273.15 K a thick CO$_2$ atmosphere forms at the ocean surface, while most of NH$_3$ is speciated and retained in the ocean as ions (NH$_4^+$ and NH$_2$COO$^-$) leading to a limited fraction of free NH$_3$ present within the ocean. This conclusion particularly contrasts with the assumption of a primordial NH$_3$-rich global ocean on Titan \citep{tobie_titans_2012}. The formation of a CO$_2$-rich atmosphere is, however, consistent with the findings of  \cite{ishimaru_oxidizing_2011}, which supports the idea that a CO$_2$-rich oxidized primordial atmosphere could have allowed the formation of N$_2$ from NH$_3$ via impact shock chemistry. 

\begin{table}[htpb]
\centering
\begin{threeparttable}

\caption{Bulk masses, partial pressures ($P_i$), and mole fractions ($x_i$) of the main volatile species at equilibrium between the primordial atmosphere and ocean at 273.15 K.} 

\begin{tabular}{@{}cccc@{}}
\hline
\hline
Molecule & Bulk mass [kg] & $P_i$ [Pa]& $x_i$\\
 & & (Atmosphere)&(Ocean)\\
\hline
H$_2$O   & 3.59$\times 10^{22}$ & 574.15 & 0.959 \\
CO$_2$   & 4.12$\times 10^{21}$ & 26.54$\times 10^{5}$  & 2.70$\times 10^{-2}$\\
NH$_3$   & 2.27$\times 10^{20}$ & 5.384$\times 10^{-3}$ & 1.26$\times 10^{-8}$\\ 
NH$_4^+$ &  --                  &  --    & 6.36$\times 10^{-3}$\\
CH$_4$   & 1.07$\times 10^{20}$ & 12.95$\times 10^{5}$  & 8.07$\times 10^{-4}$ \\ 
Ar       & 4.62$\times 10^{17}$ & 5920  & 2.71$\times 10^{-6}$\\ 
Kr       & 8.18$\times 10^{16}$ & 916.5 & 7.08$\times 10^{-7}$ \\ 
Xe       & 6.28$\times 10^{16}$ & 606.2 & 7.11$\times 10^{-7}$ \\ 
 \hline
\end{tabular}
\label{tab:hydrosphere_compo_273}
\end{threeparttable}
\end{table}
 
\begin{figure}[htpb]
\centering
\includegraphics[width=9cm]{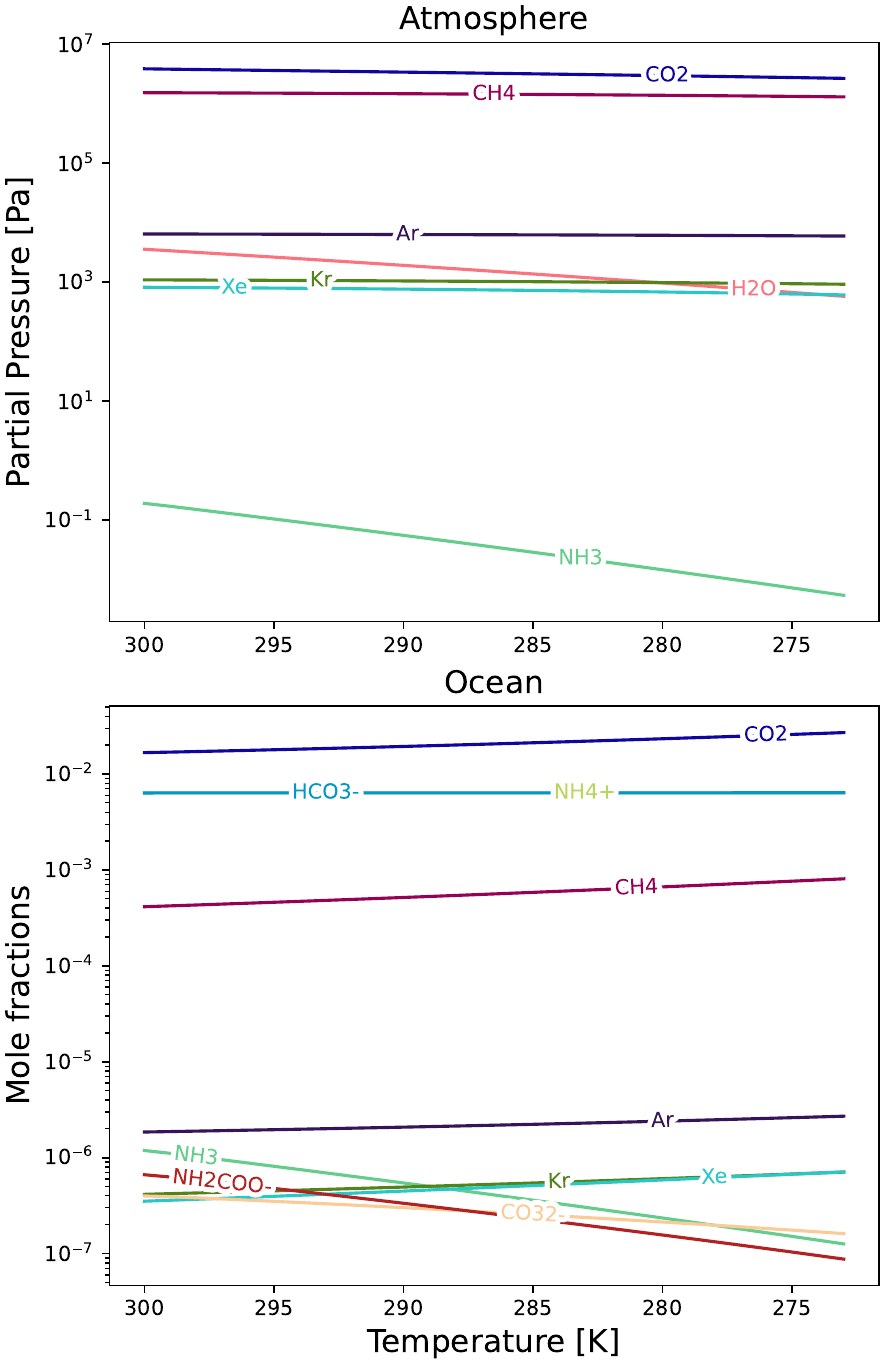}
\caption{Evolution of the composition of Titan's primordial atmosphere and ocean at equilibrium, as a function of surface temperature.} 
\label{fig:partial_pressure_T}
\end{figure}

Figure \ref{fig:partial_pressure_T} illustrates the effect of temperature on the composition of Titan's primordial atmosphere and ocean at equilibrium. As the temperature decreases from 300 K to 273.15 K, the atmosphere becomes less dense and the fraction of dissolved volatiles increases. The magnitude of this change varies with the gas under consideration. For example, in this temperature range, the partial pressure of Xe decreases by 24\%, while the abundance of Ar in the atmosphere is only slightly affected (8\%). In contrast, the evolution of the CO$_2$ and NH$_3$ abundances in the atmosphere and ocean follows a distinct trajectory, determined by their chemical equilibrium in the ocean.

\subsection{Effect of clathrate formation on the composition of the primordial atmosphere during Titan's closed-ocean phase}\label{sec:Results_clath}

\begin{figure}[htpb]
\centering
\includegraphics[width=9cm]{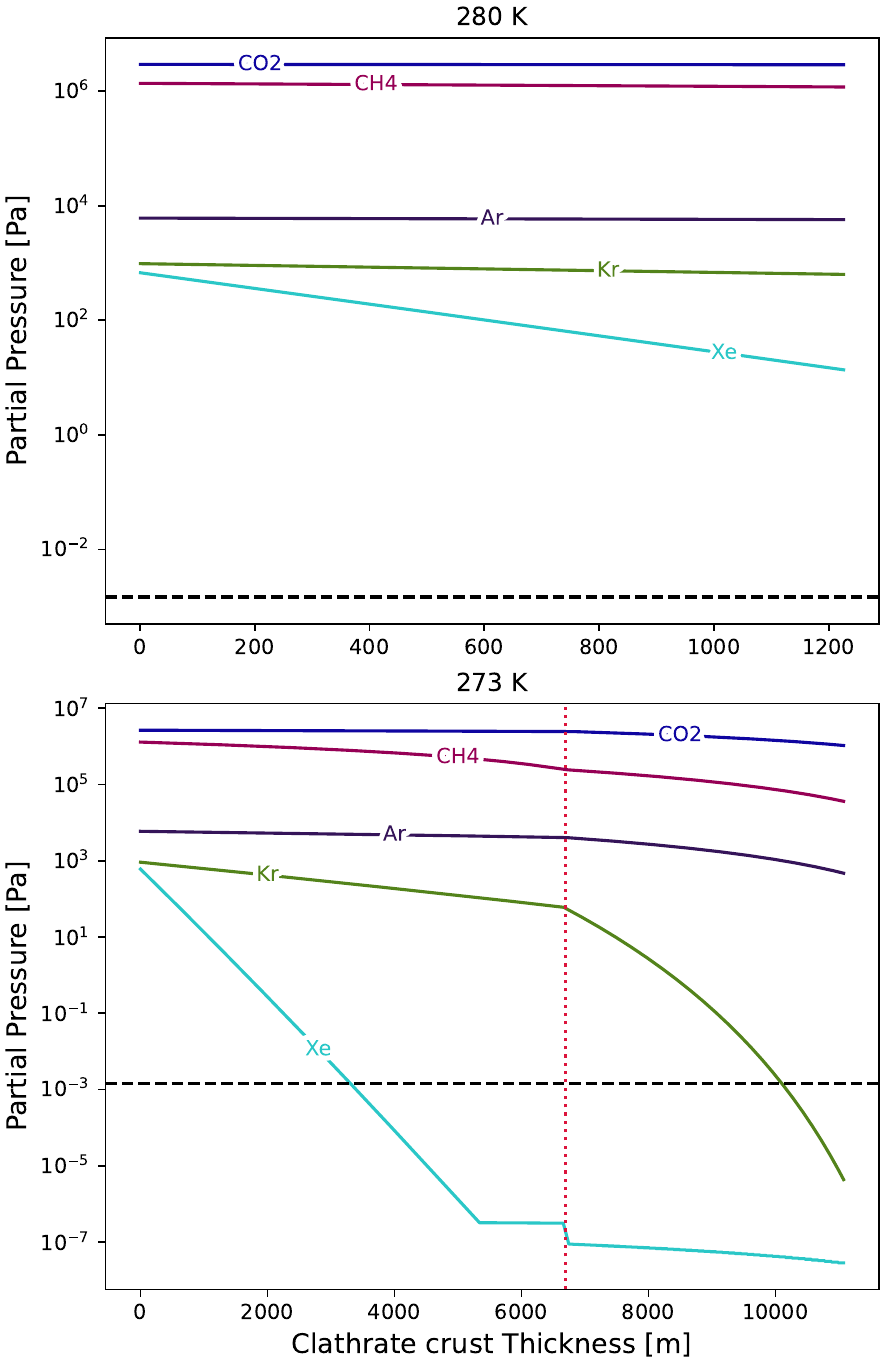}
\caption{Evolution of atmospheric abundances as a function of clathrate crust thickness at 280 and 273.15 K. The dashed red line indicates the threshold at which the clathrate composition changes from CH$_4$-rich to CO$_2$-rich. The dashed black line represents the detection limit of Huygens' GCMS.} 
\label{fig:PP_vs_crust}
\end{figure}

We focused on two stages of Titan's evolution, corresponding to a surface temperature of 280 K and 273.15 K. At those temperatures, the equilibrium total surface pressure falls within the mixed clathrate stability region, with boundaries calculated using Eq. \ref{eq:P_diss}. This boundary is reached when the surface temperature drops to 282 K, allowing clathrates to form on top of the ocean, still uncovered by the ice crust. In our model, we assumed that the clathration process halts when the surface pressure falls below the dissociation pressure of mixed clathrates. At 280 K and 273.15 K, we calculated the clathrate composition and assessed its impact on the distribution of volatiles in the atmosphere. Figure \ref{fig:PP_vs_crust} illustrates the distribution of partial pressures as a function of mixed clathrate crust thickness. The detection limit of the Huygens GCMS instrument for volatile abundances (indicated in dashed black on the figure), set at \(10^{-8}\) (corresponding to a maximum partial pressure of approximately \(1.5 \times 10^{-3}\) Pa; \citealt{niemann_abundances_2005}), serves as the threshold for determining when a gas can be considered depleted from the atmosphere. At 280 K, close to the mixed clathrate stability limit, a 1.23 km thick crust can form until the surface pressure becomes too low to allow clathrate stability. At such a low thickness, the noble gases are not trapped in significant proportions. Titan's surface must cool to reduce the dissociation pressure of the clathrates, thereby trapping more gases. At 273 K, Xe is almost completely removed from the atmosphere, becoming sufficiently rare to be undetectable by the Huygens GCMS -- when the clathrate crust reaches a thickness of 3.3 km. In contrast, trapping all of the Kr in the atmosphere requires a thicker crust, as this noble gas is less efficiently trapped than Xe. Hence, Kr is almost completely trapped in clathrates with a crust thickness of 10.31 km. 

However, Ar is much less efficiently trapped in clathrates compared to other noble gases and cannot be sufficiently captured before reaching the maximum crust thickness of 11 km at 273 K. For Ar to be completely removed from the primordial atmosphere, Titan's surface would need to cool further. Based on the total surface pressure before crust formation and the Ar abundance shown in Table \ref{tab:hydrosphere_compo_273}, our calculations suggest that complete trapping of Ar in clathrates -- whether mixed or pure -- likely did not occur during Titan's early history.

\begin{figure}
\centering
\includegraphics[width=8.5cm]{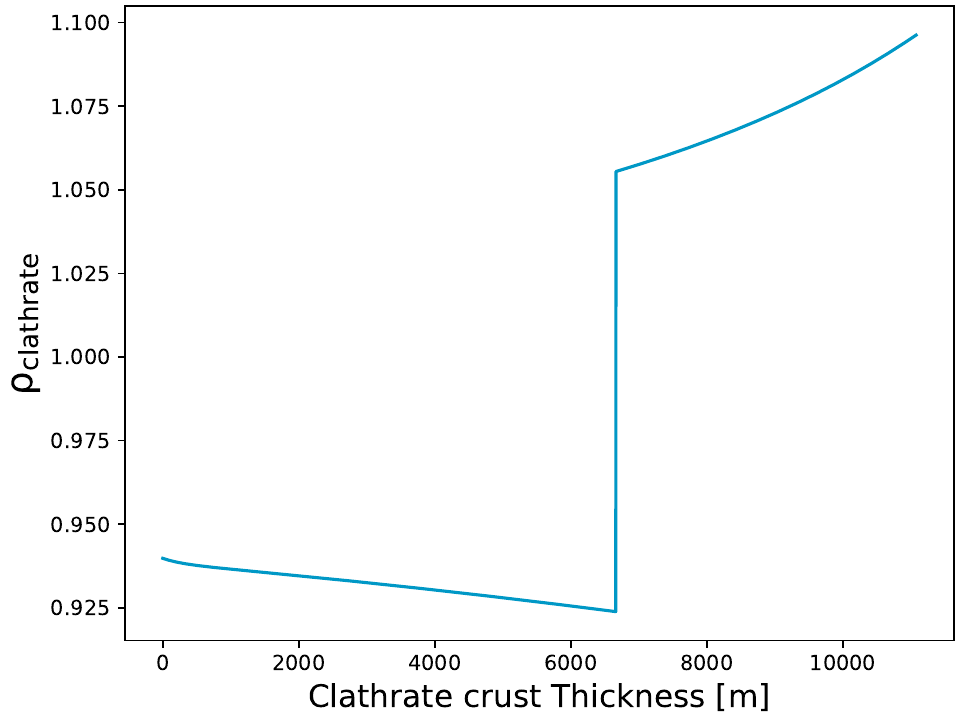}
\caption{Density of clathrates as a function of crust thickness, assuming formation at 273 K. The step for both temperatures highlights the transition from CH$_4$-rich to denser CO$_2$-rich clathrates.}
\label{fig:rho_clath}
\end{figure}

Additionally, a shift in the composition of the mixed clathrates occurs as the crust thickens and time progresses. This transition is highlighted in Fig. \ref{fig:PP_vs_crust} by the dashed red line, marking a change in the dominant gas forming the clathrate. To the left of this line, CH$_4$ is the primary gas captured in the clathrate, despite being less abundant than CO$_2$ in the atmosphere, due to thermodynamic conditions favoring its trapping. To the right of the red line, however, CH$_4$ becomes too scarce in the atmosphere, and CO$_2$ takes over as the driving force for clathrate formation. This shift in the dominant clathrate-forming gas significantly alters the overall composition of the mixed clathrates, as shown in Fig. \ref{fig:PP_vs_crust}, and consequently affects the composition of the crust over time. As shown in Fig. \ref{fig:rho_clath}, this change in composition leads to a variation in clathrate density. As the crust thickens and clathrates shift from a CH$_4$-rich to a CO$_2$-rich composition, the density of the clathrates increases, surpassing a value of 1. As a result, while CH$_4$-rich clathrates would float on the ocean, the more recent CO$_2$-rich upper layers become denser than the older, lower layers, exceeding the density of liquid water when $\rho > 1$. This density gradient could eventually lead to structural remodeling of the clathrate crust over time.

\subsection{Formation of a crustal CH$_4$-rich reservoir}

Titan's atmosphere features a $\sim$5\% CH$_4$ fraction measured at its surface by Huygens \citep{niemann_abundances_2005}. Since CH$_4$ is destroyed by UV photochemistry on 10–100 Myr timescales \citep{yung_photochemistry_1984,wilson_current_2004}, this suggests replenishment from an internal reservoir containing at least $\sim$2.8 $\times 10^{17}$ kg of CH$_4$ to maintain this level over 4.5 billion years \citep{niemann_abundances_2005}.

Assuming a comet-like primordial bulk composition, we estimated that Titan's primordial atmosphere at 300 K contained at least 9.4 $\times 10^{19}$ kg of CH$_4$, providing a sufficient initial reservoir. Once clathrates form, our calculations show that at 273.15 K, with a 11 km-thick crust, about 7.76 $\times 10^{19}$ kg of CH$_4$ would be retained. Thus, a clathrate crust just below a dozen kilometers thick could store enough CH$_4$ to serve as a long-term reservoir to maintain Titan's atmospheric composition. Based on the current mass of CH$_4$ in Titan's atmosphere, we estimate that the reservoir could replenish the atmosphere for at least 2.8 Gyr.

\section{Discussion and conclusions}

This study explores clathrate formation during Titan's primordial phase as a key factor in its present-day noble gas depletion. If Titan's volatiles originated from comet-like building blocks, a CO$_2$- and CH$_4$-rich atmosphere would have formed over a liquid water ocean depleted of free NH$_3$. In this ocean, the CO$_2$-NH$_3$ equilibrium would have converted most NH$_3$ into NH$_4^+$. Notably, if the NH$_3$ concentration in the ocean did not exceeded a few percent (which is compatible with existing constrains on comets and Enceladus data), the primordial atmosphere should be dominated by CO$_2$.
As temperatures fell to 282 K and below, clathrates could have formed a crust on the surface of Titan's ocean, incorporating Xe and Kr into a CH$_4$- and CO$_2$-rich structure and thereby accounting for their depletion in the atmosphere. In particular, if this clathrate crust reached a thickness of about 10 km at 273.15 K during Titan's early evolution, it could have efficiently sequestered Xe and Kr from the atmosphere.
However, this process could not fully remove Ar, differing from \cite{mousis_removal_2011} due to variations in atmospheric composition. We propose that Titan accreted from planetesimals that partially de-volatilized in Saturn’s sub-nebula \citep{mousis_clathration_2009}, which would explain the absence of primordial Ar. This clathrate formation may have also retained CH$_4$, contributing to its long-term atmospheric replenishment.

We emphasize that the modeling carried out in this study does not account for the clathrate formation kinetics, which is a critical aspect in this process. Specifically, our results highlighting the capture of Kr and Xe during Titan's primordial phase at 273.15 K are valid only if the clathrate crust formation occurred while the surface temperature remained at 273.15 K. Indeed, a comparison between the cooling rate of Titan and the growth rate of the clathrate crust would be necessary but remains challenging given the current state of the art. Moreover, as the crust thickens, the transport of fresh water to the surface and/or the transport of gas through the clathrate crust should be limited, affecting the capture of noble gases from the atmosphere.

The case of argon can be considered a critical aspect in our study. Argon mainly exists as $^{40}$Ar in Titan's atmosphere \citep{niemann_composition_2010}. This isotope likely originated in the rocky mantle as $^{40}$K, which subsequently decayed into $^{40}$Ar and was transported into the ocean \citep{tobie_titans_2012}. Its presence in the atmosphere suggests outgassing from the icy crust, where it may have interacted with clathrates \citep{muller-wodarg_origin_2014}. However, given the low trapping efficiency of Ar in clathrate hydrates \citep{thomas_clathrate_2007,thomas_theoretical_2008}, we propose that, in competition with other dissolved species such as CH$_4$ and CO$_2$, $^{40}$Ar most likely escaped from the crust without being incorporated into clathrates.

Finally, the bulk composition scenario and the results presented in this study raise the following questions:

\begin{itemize}
\item If a thick, primordial CO$_2$-rich atmosphere formed from comet-like building blocks shortly after Titan's accretion, what processes could account for the low present-day abundance of CO$_2$? Beyond clathrate sequestration, several plausible mechanisms include surface condensation as Titan cooled \citep{lunine_origin_2009}, early atmospheric escape \citep{kuramoto_formation_1994}, photolysis, and loss through impact-driven chemistry \citep{ishimaru_oxidizing_2011}.

\item If N$_2$ formation from NH$_3$ via shock chemistry is possible in a CO$_2$-rich atmosphere \citep{ishimaru_oxidizing_2011}, what mechanism could sustain NH$_3$ in the atmosphere, considering its expected retention in the ocean under such conditions? Possible processes include continuous outgassing from the interior, photochemical recycling, or impact-driven volatilization \citep{sekine_replacement_2011}.

\item If primordial clathrates containing CH$_4$, Ar, Kr, and Xe act as a reservoir for CH$_4$, how can CH$_4$ be released without simultaneously reintroducing noble gases — specifically, $3.74 \times 10^{16}$ kg of Xe and $5.65 \times 10^{16}$ kg of Kr — into the atmosphere? One possibility is that Xe and Kr remain sequestered in residual surface clathrates during CH$_4$ release, or are re-enclathrated shortly thereafter. This implies the existence of an additional mechanism that prevents their effective outgassing \citep{thomas_theoretical_2008, mousis_removal_2011}. The existence of such a mechanism is supported by the strong thermodynamic preference of Xe and Kr for clathrate formation compared to other gases \citep{sloan_clathrate_2008}. 
\end{itemize}

\begin{acknowledgements}
O.M. and A.B. acknowledge support from CNES. The project leading to this publication has received funding from the Excellence Initiative of Aix-Marseille Université – A*Midex, a French “Investissements d’Avenir programme” AMX-21-IET-018. This research holds as part of the project FACOM (ANR-22-CE49-0005-01 ACT) and has benefited from a funding provided by l’Agence Nationale de la Recherche (ANR) under the Generic Call for Proposals 2022.
\end{acknowledgements}

\bibliography{Bibliography}{}
\bibliographystyle{aa}

\begin{appendix}
    
\section{Description of the atmosphere-ocean equilibrium}\label{sec:L-V}
    
The model presented in this section follows the approach outlined in \cite{marounina_role_2018}, which models the H$_2$O–CO$_2$–NH$_3$ system based on previous works \citep{goppert_vaporliquid_1988, pazuki_prediction_2006, darde_modeling_2010}. Our ocean equilibrium model, developed in \cite{amsler_moulanier_role_2025} and in this study, shares several similarities with \cite{marounina_role_2018}, particularly the use of the Peng-Robinson-Gasem equation of state (EOS) for fugacity coefficients of gases and the extended UNIQUAC model for activity coefficients of dissolved species. However, several key differences distinguish our model:

\begin{itemize}
\item The standard-state fugacity of dissolved species is approximated using Henry's law constant, with an added Poynting correction factor.
\item The van der Waals one-fluid mixing rule is incorporated in the Peng-Robinson EOS to account for gas interactions.
\item Constants such as Henry's constants, $K_i$ dissociation constants, and T$_c$, P$_c$, and $\omega_c$ values are sourced from different references.
\item The activity coefficient for CH$_4$ is computed.
\item Clathrate formation at the surface of the primordial ocean is included. A key novelty of our model is simulating the impact of clathrate formation on the evolution of the primordial atmosphere. Additionally, the composition of clathrates is computed through fractional occupancies of guest molecules, using a statistical thermodynamic model.
\end{itemize}

\noindent The components of our model are outlined below.

    \subsection{Liquid-vapor equilibrium}
    
We computed the liquid-vapor equilibrium for each species starting from the equality between the vapor and liquid phases' fugacities, $f_i^V$ and $f_i^L$, respectively:

\begin{equation}
f_i^V(T,P,y_i) = f_i^L(T,P,x_i).
\label{Eq:raoult}
\end{equation}

\noindent Here, the fugacity of the species $i$ in the vapor phase, $f_i^V$, is a function of temperature $T$, pressure $P$, and mole fraction $y_i$ in the vapor phase. Likewise, the fugacity of the species $i$ in the liquid phase, $f_i^L$, is a function of $T$, $P$, and mole fraction $x_i$ in the liquid phase. 

To account for the nonideal behavior in the gas phase and the nonideal interactions between the dissolved species in the liquid phase, we introduced respectively the fugacity coefficient ($\phi$) and the activity coefficient ($\gamma$). Hence, the fugacity of each phase for an individual species $i$ can be written as

\begin{equation}
\begin{split}
f_i^V = \phi_iy_iP, \\
f_i^L = \gamma_ix_if_i^0
\end{split}
\label{Eq:fug}
.\end{equation}

\noindent In the vapor phase, the fugacity $f_i^V$ of the specie $i$ is a function of the fugacity coefficient $\phi_i$, the mole fraction $y_i$, and the total pressure $P$ in Pascal. In the liquid phase, the fugacity $f_i^L$ is a function of the activity coefficient $\gamma_i$, the mole fraction $x_i$, and the standard-state fugacity $f_i^0$.

Then, we were able to compute the partitioning of volatiles between the atmosphere and the ocean, taking the nonideal behavior of both phases into account, by substituting these expressions into Eq. \eqref{Eq:raoult}. The standard-state fugacity ($f_i^0$) of H$_2$O is set to the saturation pressure of water, $P_{H_2O}^s(T)$, which is calculated using Antoine's equation \citep{stull_vapor_1947}: 

\begin{equation}
    \log_{10}(P^{s}_{H_2O}) = A - \cfrac{B}{T+C},
    \label{Eq:antoine}
\end{equation}

\noindent with $A=4.6543$, $B=1435.264$, $C=-64.848$, \textit{T} in K and \textit{P} in bar.

Equation \eqref{Eq:fug} can then be rewritten as\begin{equation}
\begin{split}    
&\phi_{H_2O} \; P\; y_{H_2O} = \gamma_{H_2O}\; x_{H_2O}\; P_{H_2O}^s(T) \\
&\phi_{i} P y_{i} = \gamma_{i}\; x_{i}\; H_{H_2O,i}(T)\exp\left(\cfrac{v_{i,H_2O}^{\infty}(P-P_{H_2O}^s)}{RT}\right),
\end{split}
\label{Eq:eq_LV_equation}
\end{equation}

\noindent where the standard-state fugacity $f_i^0$ of dissolved species $i$ is approximated as the Henry's law constant, corrected by a Poynting factor \citep{kawazuishi_correlation_1987,darde_modeling_2010}. The Henry's law constants and their temperature dependences are reported {in Table \ref{tab:henrys_value}} and were taken from \cite{kawazuishi_correlation_1987} for CO$_2$ and NH$_3$, and from \cite{warneck_atmospheric_2012} for the other species. 

\begin{table*}[htpb]
\centering
\begin{threeparttable}[b]

\caption{Henry's constants' interpolation formulas of gases in water.}
\begin{tabular}{@{}ccc@{}}
\hline
\hline
\smallskip
Molecule & Henry's constant formula & Unit\\
\hline
\smallskip
CO$_2$\tnote{{\bf a}}   & ln(H$_i$)= $196.876 - 9624.4/T + 0.01441T - 28.749\ln{T}$   & MPa.kg.mol$^{-1}$      \\
NH$_3$\tnote{{\bf a}}   & ln(H$_i$)= $3.932 - 1879.02/T - 355134.1T^2$   & MPa.kg.mol$^{-1}$  \\ 
CH$_4$\tnote{{\bf b}}   & ln(H$_i$)= $-211.28 + 10447.9/T + 29.780\ln{T}$ & mol.dm$^{-3}$.atm$^{-1}$      \\ 
Ar\tnote{{\bf b}}       & ln(H$_i$)= $-146.40 + 7479.3/T + 20.140\ln{T}$  & mol.dm$^{-3}$.atm$^{-1}$      \\ 
Kr\tnote{{\bf b}}       & ln(H$_i$)= $-174.52 + 9101.7/T + 24.221\ln{T}$  & mol.dm$^{-3}$.atm$^{-1}$      \\ 
Xe\tnote{{\bf b}}       & ln(H$_i$)= $-197.21 + 10521.0/T + 27.466\ln{T}$ & mol.dm$^{-3}$.atm$^{-1}$      \\ 
\hline
\end{tabular}
\label{tab:henrys_value}
\begin{tablenotes}
\item [{\bf a}] \cite{rumpf_experimental_1993,rumpf_solubility_1993}.
\item [{\bf b}] \cite{warneck_atmospheric_2012}.
\end{tablenotes}
\end{threeparttable}
\end{table*}

The molar volume of the species $i$ at infinite dilution, $v_{i,H_2O}^{\infty}$, is extrapolated from \cite{rumpf_experimental_1993,rumpf_solubility_1993} and \cite{garcia_density_2001} for CO$_2$ and NH$_3$. For the other species, $v_{i,H_2O}^{\infty}$ is set to 0. $R$ is the ideal gas constant. 

We computed the fugacity coefficient ($\phi_i$) for each species using the Peng-Robinson EOS \citep[][]{peng_new_1976} and the van der Waals one-fluid mixing rule. 
The molar volume of the mixture $v$ was derived from the Peng-Robinson EOS: 

\begin{equation}
P = \cfrac{RT}{v-b} - \cfrac{a\alpha}{v^2 +2vb - b^2},
\label{eq:EOSPG}
\end{equation}

\noindent where 

\begin{equation}
\begin{split}  
&a = \cfrac{0.457235R^2T_c^2}{P_c},\\
&b = \cfrac{0.077796RT_c}{P_c},\\
&\alpha = \left(1+\kappa\left(1-\sqrt{\cfrac{T}{T_c}}\right)\right)^2,\\
{\rm and}~&\kappa = 0.37464 + 1.54226\omega - 0.2699\omega^2.
\end{split}
\label{eq:parametersEOS}
\end{equation}

\noindent Here, $T_c$ is the critical temperature, $P_c$ is the critical pressure, and $\omega$ is the acentric factor of each volatile. Their values are shown in Table \ref{tab:critical_values}.

The mixture parameters $A_{mix}$ and $B_{mix}$ required for the Peng-Robinson EOS and the van der Waals one-fluid mixing rule were computed using the following equations:

\begin{equation}
\begin{split}
&A_i=a_i\times\alpha_i,\\
&A_{ij} = (1-k_{ij})\sqrt{A_iA_j},\\
&A_{mix} = \sum_{i}^{N}\sum_{j}^{N} y_iy_jA_{ij},\\
{\rm and}~&B_{mix} = \sum_{i}^{N}y_ib_i,
\end{split}
\end{equation}

\noindent where $N$ is the total number of species. $\alpha_i$, $a_i$ and $b_i$ are computed for each species $i$ using Eq. \ref{eq:parametersEOS}. In order to better capture the nonideal behavior of the volatile mixtures in the atmosphere and ocean, binary interaction coefficients $k_{ij}$ are used to further account for the nonideal interactions between the volatile species, following the approach described in \cite{gasem_modified_2001}. If they are not available in the literature, they are set to 0. These binary interaction coefficients are determined empirically and taken from the literature (\cite{dhima_solubility_1999} for H$_2$O-CO$_2$, and \cite{vrabec_molecular_2009} for the other couples).
After calculating the molar volume $v$ from the Peng-Robinson EOS (Eq. \ref{eq:EOSPG}), the fugacity coefficient $\phi_i$ of a species $i$ can be derived from Eq. \ref{eq:EOSPG}:

\begin{equation}
\begin{split}
    \ln(\phi_i(T,v,y)) &= \cfrac{b_i}{B_{mix}}(Z-1) - \ln\left(\cfrac{P}{RT}(v-B_{mix})\right)+ \\
    &\cfrac{A_{mix}}{2\sqrt{2}RTB_{mix}}\ln \left (\cfrac{v + B_{mix}2.414}{v- B_{mix}0.414}\right) 
\end{split}
\end{equation}

\noindent with, $Z = \cfrac{Pv}{RT}.$

The activity coefficients of H$_2$O, CO$_2$ and NH$_3$ are computed with the extended UNIQUAC model \citep{thomsen_modeling_1999}. The activity coefficient of CH$_4$ is taken from \cite{kvamme_small_2021}. For the other neutral species, we set $\gamma_i=1$.

\begin{table}[htpb]
\centering
\caption{$T_c$, $P_c$, and $\omega$ values used in the model \citep{reid_properties_1987}.}
\begin{tabular}{@{}cccc@{}}
\hline
\hline
\smallskip
Molecule & $T_c$ & $P_c$ & $\omega$ \\
\hline
\smallskip
H$_2$O   & 647.3    & 220.6      & 0.3434 \\
CO$_2$   & 304.19   & 73.83      & 0.224  \\
NH$_3$   & 405.6    & 112.8      & 0.25    \\ 
CH$_4$   & 190.4    & 46.0       & 0.011   \\ 
Ar       & 150.8    & 48.7       & 0.001   \\ 
Kr       & 209.4    & 55.0       & 0.005   \\ 
Xe       & 289.7    & 58.4       & 0.008   \\ 
\hline
\end{tabular}
\label{tab:critical_values}
\end{table}

    \subsection{Chemical equilibrium}

The liquid-vapor equilibrium is coupled with the chemical equilibrium taking place within the aqueous H$_2$O-CO$_2$-NH$_3$ system:  
\begin{align}
        H_2O &\rightleftharpoons{} H^+ + OH^-, \\
        \label{eq:carbonate}
        CO_2 + H_2O &\rightleftharpoons{} HCO_3^- + H^+, \\
        \label{eq:ammonium}
        NH_3 + H_2O &\rightleftharpoons{} NH_4^+ + OH^-, \\
        \label{eq:bicarbonate}
        HCO_3^- &\rightleftharpoons{} CO_3^{2-} + H^+, \\
        \label{eq:carbamate}
        HCO_3^- + NH_3 &\rightleftharpoons{} NH_2COO^- + H_2O.    
\end{align}
The equations describing the chemical equilibrium within the H$_2$O-CO$_2$-NH$_3$ system are derived from the dissociation reactions of these species:

\begin{align}
    K_{H_2O} &= \cfrac{m_{OH^-}m_{H^+}}{x_{H_2O}} \; \cfrac{\gamma_{OH^-}\gamma_{H^+}}{\gamma_{H_2O}}, \\
    K_{CO_2} &= \cfrac{m_{HCO_3^-}m_{H^+}}{x_{H_2O}m_{CO_2}} \; \cfrac{\gamma_{HCO_3^-}\gamma_{H^+}}{\gamma_{H_2O}\gamma_{CO_2}}, \\
    K_{NH_3} &= \cfrac{m_{NH_4^{+}}m_{OH^-}}{x_{H_2O}m_{NH_3}} \; \cfrac{\gamma_{NH_4^+}\gamma_{OH^-}}{\gamma_{H_2O}\gamma_{NH_3}}, \\
    K_{HCO_3^-} &= \cfrac{m_{CO_3^{2-}}m_{H^+}}{m_{HCO_3^-}} \; \cfrac{\gamma_{CO_3^{2-}}\gamma_{H^+}}{\gamma_{HCO_3^{-}}}, \\
    K_{NH_2COO^-} &= \cfrac{m_{NH_2COO^-}x_{H_2O}}{m_{NH_3}m_{HCO_3^-}} \; \cfrac{\gamma_{NH_2COO^-}\gamma_{H_2O}}{\gamma_{NH_3}\gamma_{HCO_3^-}},
\end{align}

\noindent where $K_i(T)$ is the dissociation constant of each reaction taken from \cite{kawazuishi_correlation_1987}, $m_i$ (in mol/kg) is the molality of the aqueous species, $x_{H_2O}$ the mole fraction of water, and  $\gamma_i$ the activity coefficient of the species $i$.

To compute the activity coefficients involved in the chemical equilibrium equations, we used the extended UNIQUAC model, first introduced in \cite{thomsen_modeling_1999} and the parameters from \cite{darde_modeling_2010}. This model calculates the activity coefficient using combinatorial, residual, and electrostatic terms, with the  electrostatic terms based on the extended Debye-Hückel law. It is well established for studying this specific chemical equilibrium \citep{thomsen_modeling_1999,thomsen_modeling_2005,darde_modeling_2010}.
Finally, we could compute the whole set of variables using the mass and charge balance equations derived as follows: 

\begin{align}
&m_{CO_{{2}{tot}}} = m_{CO_{2}} + m_{HCO_{3}^-} + m_{CO_{3}^{2-}} + m_{NH_{2}COO^{-}}, \\
&m_{NH_{{3}{tot}}} = m_{NH_{3}} + m_{NH_{4}^{+}} + m_{NH_{2}COO^{-}}, \\
&m_{H^{+}} + m_{NH_{4}^{+}} = m_{OH^{-}} + m_{HCO_{3}^{-}} + 2m_{CO_{3}^{2-}} + m_{NH_{2}COO^{-}},
\label{eq:balance}
\end{align}

\noindent where $m_{CO_{{2}{tot}}}$ and $m_{NH_{{3}{tot}}}$ are the total molalities of CO$_2$ and NH$_3$ incorporated into the liquid phase.

\section{Clathrate formation model}\label{sec:app_clath}
\subsection{The statistical thermodynamic model}

The method described in this section, detailed in \cite{mousis_abundances_2013}, uses classical statistical mechanics to describe the macroscopic thermodynamic properties of clathrates to the molecular structure and interaction energies. It is based on the original model of \cite{interscience_clathrate_1959}, which assumes that trapping of guest molecules into cages corresponds to the three-dimensional generalization of ideal localized adsorption. It is based on four hypotheses: 
\begin{itemize}
    \item The contribution of the host molecules to the free energy is independent of the clathrate occupancy. This implies that the guest species do not distort the cages. 
    \item The cages are singly occupied, and guest molecules rotate freely within the cage. 
    \item Guest molecules do not interact with each other. 
    \item Classical statistics is valid, meaning that quantum effects are negligible. 
\end{itemize}

Hence, the fractional occupancy of a guest molecule $i$ for a given type of cage $q$ (small or large) in a clathrate structure can be written as \begin{equation}
    y_{i,q} = \cfrac{C_{i,q}f_{hydro,i}}{1 + \sum_J  C_{J,q}f_{hydro,J}}
    \label{eq:y_i}
,\end{equation}
where the sum in the denominator includes all the species present in the gas. C$_{i,q}$ is the Langmuir constant of the species $i$ in the cage of type $q$, $f_{hydro,i}$ is the fugacity of the guest species $i$ computed at the total gas pressure (see Appendix \ref{sec:app_fuga}). 

The Langmuir constant C$_{i,q}$ depends on the strength of the interaction between each guest species and each type of cage. It can be calculated by integrating the molecular potential energy within the cavity as  
\begin{equation}
     C_{i,q} = \frac{4\pi}{k_BT} \int_0^{R_c} exp(-\frac{w_{i,q}(r)}{k_BT})r^2dr
,\end{equation}
where $R_c$ represents the radius of the cavity assumed to be spherical, $k_B$ its the Boltzmann constant, and $w_{i,q}(r)$ is the spherically averaged potential (here Kiahara or Lennard-Jones potential) representing the interactions between the guest molecules $i$ and the water molecules forming the surrounding cage $q$. This potential $w_{i,q}(r)$ can be written for a spherical guest molecule as  \citep{mckoy_theory_1963}  
\begin{equation}
    w(r) = 2z\epsilon \left[ \frac{\sigma^{12}}{R_c^{11}r} \left(\delta^{10}(r) + \frac{a}{R_c}\delta^{11}(r) \right) - \frac{\sigma^6}{R_c^5r} \left(\delta^4(r) + \frac{a}{R_c}\delta^5(r) \right) \right]
    \label{eq:w}
,\end{equation}
with the mathematical function $\delta^N(r)$ in the form 
\begin{equation}
        \delta^N(r) = \frac{1}{N} \left[ \left( 1 - \frac{r}{R_c} - \frac{a}{R_c} \right)^{-N} - \left(1 + \frac{r}{R_c} - \frac{a}{R_c} \right)^{-N} \right]
.\end{equation}
In Eq. \ref{eq:w}, $z$ is the coordination number of the cell. Parameters $z$ and $R_c$ depend on the cage type $q$ (small or large) and on the clathrate structure (I or II). Their values are given in Table \ref{tab:parameters_cavity}. The intermolecular parameters $a$, $\sigma$ and $\epsilon$ describing  the guest molecule $i$-water (i-W) interactions in the form of a Kihara or Lennard-Jones potential are listed in Table \ref{tab:potentials}. 

\begin{table}
\centering
\begin{threeparttable}
\caption{Parameters for the clathrate cavities.}

\begin{tabular}{lcccc}
\hline
\hline
Clathrate structure&  \multicolumn{2}{c}{\textit{I}}&  \multicolumn{2}{c}{\textit{II}}\\
\hline
$N_w$&  \multicolumn{2}{c}{$46$}&  \multicolumn{2}{c}{$136$}\\
$V_c$(\text{\AA}$^3$)&  \multicolumn{2}{c}{$12.0^3$}&  \multicolumn{2}{c}{$17.3^3$}\\
Cavity type&  small&  large&  small& large\\
$R_c$ (\text{\AA}$^3$)&  3.975&  4.300&  3.910& 4.730\\
$b$&  2&  6&  16& 8\\
$z$&  20&  24&  20& 28\\
\hline
 & & & &\\
\end{tabular}
\label{tab:parameters_cavity}
\begin{tablenotes}
    \item $R_c$ is the radius of the cavity (values taken from \citealt{parrish_dissociation_1972}). $b$ represents the number of small ($b_s$) or large ($b_l$) cages per unit cell of a given structure (I or II) with volume $V_c$ \citep{sloan_clathrate_2008}. $z$ is the coordination number in a cavity. 
\end{tablenotes}
\end{threeparttable}
\end{table}

\begin{table*}
\centering
\begin{threeparttable}
\caption{Parameters for the Kihara and Lennard-Jones potentials.}

\begin{tabular}{lcccc}
\hline
\hline
Molecule&  $\sigma_{i-w}$&$\epsilon_{i-w}$&  $a_{i-w}$&Reference\\
\hline
CO$_2$&  2.97638&175.405&  0.6805& \cite{sloan_clathrate_2008}\\
CH$_4$&  3.14393&155.593&  0.3834& \cite{sloan_clathrate_2008}\\
Ar&  2.9434&  170.50&  0.184& \cite{parrish_dissociation_1972}\\
Kr&  2.9739&  198.34&  0.230& \cite{parrish_dissociation_1972}\\
Xe&  3.32968&  193.708&  0.2357& \cite{sloan_clathrate_2008}\\
\hline
\end{tabular}
\label{tab:potentials}
\begin{tablenotes}
    \item $\sigma_{i-w}$ is the Lennard-Jones diameter,  $\epsilon_{i-w}$  is the depth of the potential well, and $a_{i-w}$ is the radius of the impenetrable core, for the guests-water pairs. 
\end{tablenotes}
\end{threeparttable}
\end{table*}

Finally, the mole fraction $x_i^{clath}$ of a guest molecule $i$ in a clathrate of a given structure (I or II) can be computed as 
\begin{equation}
    x_i^{clath} = \cfrac{b_sy_{i,s} + b_ly_{i,l}}{b_s \sum_J y_{J,s} + b_l \sum_J y_{J,l}}
,\end{equation}
where $b_s$ and $b_l$ are the number of small and large cages per unit cell, respectively, for the clathrate structure we considered. The mole fractions of the enclathrated species is normalized to 1.

\subsection{Calculation of the fugacities}\label{sec:app_fuga}

To determine $f_{hydro,i}$, we first calculated the specific volume ($v$) of the considered mixture via the Redlich-Kwong EOS \citep{redlich_thermodynamics_1949}:
\begin{equation}
    P = \cfrac{RT}{v-b} - \cfrac{a}{\sqrt{T}v(v+b)}
\end{equation}
with 
\begin{empheq}{align}
    & a = 0.4278\frac{R^2T_c^{2.5}}{P_c},
    & b = 0.08664\frac{RT_c}{P_c}
\end{empheq}

\noindent where $R$ is the perfect gas constant, $T$ the ambient temperature, $T_c$ and $P_c$ the critical temperature and pressure of the substance $i$ Tab.\ref{tab:critical_values}. In this case, $a$ and $b$ account for a pure substance. For mixtures, we took the average properties of the mixture  into account  by using the mixing rules

\begin{empheq}{align}
        &a_m = \sum_i \sum_j y_iy_ja_{ij}
        &{\rm with}~ a_{ij} = \sqrt{a_ia_j},
\end{empheq}
\[b_m = \sum_i y_ib_i,\]

\noindent where $y_i$ the molar fraction of the gas in the atmosphere, normalized for the enclathrated species to 1. 
Then, we defined the mixture pressure $P_m$ dissolved in water, which corresponds to the sum of the individual dissolved pressure $P_i$. This pressure is different than the partial pressure of the gas in the atmosphere. It is expressed as a function of Henry's law coefficient as  
\begin{equation}
    P_i = H_{hydro,i} \times y_i
,\end{equation}
where $H_{hydro,i}$ is the Henry's law constant calculated for specie $i$ at the surface (hydrostatic) pressure $P_{tot}$ and temperature $T$. It is calculated as 
\begin{equation}
    \text{ln}(\frac{H_{hydro,i}}{H_{1 atm,i}}) = \frac{ \bar{V_i}}{RT}(P_{tot} - 1.013\times 10^5)
,\end{equation}
where $H_{1atm,i}$ is the Henry's constant of specie $i$ calculated at 1 atmosphere and 0°C and $\bar{V_i}$ is the partial molar volume of specie $i$. 
The fugacity of the mixture at 1 atm pressure ($f_{1atm,m}$) can be computed as \citep{redlich_thermodynamics_1949}
\begin{equation}
\begin{split}
        \text{ln}f_{1atm,m} = \frac{b_m}{v_m - b_m} + \text{ln}\frac{RT}{v_m - b_m}  - \\
        \frac{a_m}{RT^{3/2}} \left(\frac{1}{v_m + b_m} + \frac{1}{b_m}\text{ln}\frac{(v_m + b_m)}{v_m} \right)
\end{split}
.\end{equation}

The fugacity of the mixture calculated at the surface pressure $f_{hydro,m}$ is related to $f_{1atm,m}$ via \citep{kaplan_nature_1974}  
\begin{equation}
    \text{ln}\frac{f_{hydro,m}}{f_{1atm,m}} = -n \text{ln}a_w - \cfrac{P_{tot} - 1.013\times10^5}{RT}(vV_{H_2O} - V_{c,m})
,\end{equation}
where $a_w$ is the activity of water relative to pure liquid water ($\sim 1$ here), $n$ is the moles of water per mole of species $i$ in the clathrate ($n \sim 6$), $V_{H_2O}$ is the molar volume of liquid water ($1.8 \times 10^{-5}$ m$^3$) and $V_{c,m}$ is the volume of clathrate that contains 1 mol of substance $m$. 
Finally, the fugacity coefficient $\phi$ of the mixture can be computed from 
\begin{equation}
    f_{hydro,i} = \phi \times P_i
.\end{equation}
The values of $T_c$, $P_c$, $\bar{V_i}$,  $H_{1atm,i}$ and $f_{1atm,i}$ used for each species $i$ are taken from \cite{mousis_abundances_2013} Table 1. 

\subsection{Calculation of clathrate density}\label{sec:density}

From the calculation of the fractional occupancies ($y_{i,q}$) in both small and larges cages, it is possible to derive the clathrate density based upon a unit crystal \citep{sloan_clathrate_2008}: 
\begin{equation}
    \rho = \cfrac{N_wM_{H_2O} + \sum^C_{q=1}\sum^N_{i=1}y_{i,q}b_qM_i}{N_{Ava}V_{cell}}
,\end{equation}
where $N_W$ is the number of water molecules per unit cell, $N_{Ana}$ is Avogadro's number ($6.023 \times 10^{23}$), $M_i$ the molecular weight of molecule $i$, $y_{i,q}$ is the fractional occupancy of molecule $i$ in type $q$ cavity (small or large), $b_q$ is the number of type $q$ cavities per water molecules in unit cell , $V_{cell}$ is the volume of an unit cell, $N$ is the number of cavity type per unit cell (in this case 2) and $C$ the number of component trapped in hydrate phase. $N_w$, $b_q$ and V$_{cell}$ values are given in Table \ref{tab:parameters_cavity} for each structure (I and II). 
\end{appendix}

\end{document}